\documentclass[twocolumn]{aastex631}
\usepackage{amsmath}
\usepackage{booktabs}

\submitjournal{ApJ}

\shorttitle{Rings Around HIP-41378 f}
\shortauthors{Lu et al.}
\graphicspath{{./}{figures/}}

\begin{document}

\title{The Dynamical History of HIP-41378 f - Oblique Exorings Masquerading as a Puffy Planet}

\correspondingauthor{Tiger Lu}
\email{tiger.lu@yale.edu}

\author[0000-0003-0834-8645]{Tiger Lu}
\affiliation{Department of Astronomy, Yale University, 219 Prospect Street, New Haven, CT 06511, USA}

\author[0000-0001-8308-0808]{Gongjie Li}
\affiliation{Center for Relativistic Astrophysics, School of Physics, Georgia Institute of Technology, Atlanta GA 30332, USA}

\author[0000-0002-9544-0118]{Ben Cassese}
\affiliation{Dept. of Astronomy, Columbia University, 550 W 120th Street, New York NY 10027, USA}

\author[0000-0001-5466-4628]{D.N.C Lin}
\affiliation{Dept. of Astronomy and Astrophysics, University of California, Santa Cruz, CA 95064, USA}

\begin{abstract}
The super-puff HIP-41378 f represents a fascinating puzzle due to its anomalously low density on a far-out orbit in contrast with other known super-puffs. In this work, we explore the hypothesis that HIP-41378 f is not in fact a low-density planet, but rather hosts an opaque ring system. We analyze the dynamical history of the system, and show that convergent migration is necessary to explain the system's long-term stability. We then show that this same migration process plausibly captures HIP-41378 f into spin-orbit resonance and excites the planetary obliquity to high values. This tilts the surrounding ring and is a plausible explanation for the large transit depth. In the end, we also briefly comment on the likelihood of other super-puff planets being in high-obliquity states. We show that the existence of a tilted extensive ring around a high obliquity planet can serve as an explanation for puffy planets, particularly in multi-planetary systems at far distances from their host stars.

\end{abstract}
%\keywords{Classical Novae (251) --- Ultraviolet astronomy(1736) --- History of astronomy(1868) --- Interdisciplinary astronomy(804)}

\section{Introduction} \label{sec:intro}
In recent years, a class of planets with extremely low densities has emerged. These planets, the so-called ``super-puffs" \citep{lee_chiang_2016}, have masses a few times greater than Earth's and radii in excess of Neptune's, resulting in densities of $\rho < 0.3 \: \mathrm{g/cm}^{-3}$. Standard thermal evolution models \citep[e.g.][]{rogers_2011, batygin_stevenson_2013, lopez_fortney_2014} have been seriously challenged to reproduce these super-puffs, often requiring envelope mass fractions well in excess of $20\%$ which lies in tension with standard core accretion models of planet formation. 

A number of explanations have been put forth to explain these anomalous planets. \cite{lee_chiang_2016} posit that super-puffs form in the outskirts of the circumstellar disk where it is easier to accrete large amounts of material, and migrated to their present-day locations. Tidal inflation is another possibility -- \cite{millholland_2019_tides, millholland_2020} demonstrated that tidal heating is sufficient to inflate planets with more standard envelope mass-fractions of $1-10\%$. The question of planetary inflation is perhaps best studied in hot Jupiters, where the hot Jupiter radius anomaly remains an open question \citep[e.g.][]{fortney_2021}. While the exact mechanism responsible for this inflation is uncertain \citep[e.g.][]{fabrycky_2007, batygin_2010, leconte_2010, tremblin_2017}, there is a clear trend linking radius inflation and effective temperature \citep{laughlin_2011} which indicates inflation relies on the super-puff in question orbiting close-in to its host star. Most discovered super-puffs do indeed have the requisite close-in orbit. However, this trend was broken with the discovery of the HIP-41378 system by \cite{vanderburg_2016}. The outermost planet, HIP-41378 f, was found to be a super-puff by \cite{santerne_2019}, and this presents an intriguing mystery. HIP-41378 f exhibits an extremely low density of $\rho \sim 0.09 \: \mathrm{g/cm}^{-3}$ yet orbits its host star with a period of $542$ days, well beyond the orbit of Earth. The relevant physical and orbital parameters of the system are provided in Table \ref{tab:params}. \cite{belkovski_2022} showed that no reasonable interior structure consistent with standard formation theory is capable of reproducing the anomalous density of HIP-41378 f, and conclude that the planet's low density is incompatible with the traditional methods of inflation.

\begin{table*}[]
    \centering
    \begin{tabular}{lccccccc}
    \toprule
     \textbf{Planet} & Mass ($M_\mathrm{E}$)& Radius $(R_\mathrm{E})$ & $\rho$ (g/cm$^3$) & Orbital Period (days) & Semimajor Axis (AU) & Eccentricity & Inclination $(^\circ)$\\
     \toprule
      \textbf{b}  & $6.89 \pm 0.88$ & $2.595 \pm 0.036$ & $2.17 \pm 0.28$ & $15.57208 \pm 2 \times 10^{-5}$ & $0.1283 \pm 1.5 \times 10^{-3}$ & $0.07 \pm 0.06$ & $88.75 \pm 0.13$\\
      
      \textbf{c}  & $4.4 \pm 1.1$ & $2.727 \pm 0.060$ & $1.19 \pm 0.30$ & $31.70603 \pm 6 \times 10^{-5}$ & $0.2061 \pm 2.4 \times 10^{-3}$ & $0.04^{0.04}_{-0.03}$ & $88.47^{+0.035}_{-0.061}$\\

      \textbf{d}  & $<4.6$ & $3.54 \pm 0.06$ & $<0.56$ & $278.3618 \pm 5 \times 10^{-4}$ & $0.88 \pm 0.01$ & $0.06 \pm 0.06$ & $89.80 \pm 0.02$\\

      \textbf{e}  & $12 \pm 5$ & $4.92 \pm 0.09$ & $0.55 \pm 0.23$ & $369 \pm 10$ & $1.06^{+0.03}_{-0.02}$ & $0.14 \pm 0.09$ & $89.84^{+0.07}_{-0.03}$\\

      \textbf{f}  & $12 \pm 3$ & $9.2 \pm 0.1$ & $0.09 \pm 0.023$ & $542.07975 \pm 1.4 \times 10^{-4}$ & $1.37 \pm 0.02$ & $0.004^{+0.009}_{-0.003}$ & $89.971^{+0.01}_{-0.008}$\\
      \bottomrule
    \end{tabular}
    \caption{Relevant physical and orbital parameters of the planets in the HIP-41378 f system, as reported in \cite{santerne_2019}.}
    \label{tab:params}
\end{table*}

Given the difficulty of creating such a low-density planet far from its host star, an appealing explanation is that the planet is not actually an extremely low density planet and only appears to be due to some obscuring effect masquerading as a large radius in the transit lightcurve. One explanation is atmospheric hazes and a dusty outflowing atmosphere \citep{wang_dai_2019}. The most popular hypothesis, which we explore in this work, is the existence of an opaque planetary ring system around the planet \citep{akinsanmi_2020}. While there has to date not been a direct detection of a exoplanetary ring system, there is some evidence to support this hypothesis -- \cite{ohno_fortney_2022} demonstrate that the presence of planetary rings or hazes results in a featureless transmission spectrum, which was observed by \cite{alam_2022}. They also conclude that planetary rings are long-term stable around planets only if $T_\mathrm{eq} < 300$ K, which rules out the majority of super-puffs but is consistent with HIP-41378 f's equilibrium temperature of 294 K.

A crucial piece of the exoring hypothesis is planetary obliquity -- for a ring system to be visible in the transit lightcurve the planet must have a nonzero planetary obliquity, lest the rings be viewed edge-on and therefore contribute nothing to the transit depth \citep[e.g.][]{barnes_fortney_2004}. \cite{saillenfest_hip} proposed a formation mechanism involving a migrating exomoon which consistently explains the planetary obliquity and formation of the ring system. \cite{harada_2023} showed that such large moons are tidally and dynamically stable around HIP-41378 f. In this work we explore an alternative dynamical history of the HIP-41378 system. We show that the system was likely delivered to its present-day orbital configuration via convergent migration, and that excited planetary obliquity sufficient to reproduce the observed transit depth is a natural consequence of this migration. The structure of this paper is as follows. In Section \ref{sec:architecture} we discuss the architecture of the HIP-41378 system, and show that it is unstable unless convergent migration occurred in its past. In Section \ref{sec:rings} we describe the extent and configuration of a realistic planetary ring system around HIP-41378 f, and the resulting impact on the transit lightcurve as a function of ring composition and orientation. In Section \ref{sec:spin_orbit_resonance} we provide a brief background on the mechanism of secular spin-orbit resonance, and in Section \ref{sec:hip_sims} we use \textit{N}-body simulations to demonstrate that such a process could have reasonably excited the obliquity of HIP-41378 f. We explore the possibility of the other super-puffs being in high-obliquity states in Section \ref{sec:others}. We conclude in Section \ref{sec:conclusion}.

\section{Architecture and Stability of the HIP-41378 System}
\label{sec:architecture}
HIP-41378 is a roughly solar-mass F-type star which hosts five transiting planets. The relevant orbital elements and physical parameters are given by \cite{santerne_2019} and reproduced in Table \ref{tab:params}. We note that recently \cite{sulis_2024} targeted HIP-41378 d with CHEOPS during the predicted transit timing and did not detect a transit, which casts some doubt on the original orbital parameters put forth by \cite{santerne_2019}. They propose either a misidentified period or a large transit-timing variation to explain the missing transit. We assume the second explanation for now, and proceed with the originally determined parameters.

Most significantly, the outer three planets are very closely packed. The mutual separation of multi-planet systems in commonly parameterized by the mutual Hill radius:

\begin{equation}
    R_\mathrm{H} = \frac{a_{i} + a_{i-1}}{2} \left(\frac{m_{i} + m_{i-1}}{3 M_*}\right)^{1/3}.
\end{equation}
The mutual separations between planets $d$, $e$ and planets $e$, $f$ are $7.6 \ R_\mathrm{H}$ and $9.3 \ R_\mathrm{H}$, respectively. Systems this compact are typically dynamically unstable on timescales of around $10^7$ orbits, or a few Myr \citep[e.g.][]{obertas_2017, gillon_trappist}. Given the system age of $3.1 \pm 0.6$ Gyr \citep{santerne_2019} it is extremely unlikely that we are observing the system in this small stable window. Indeed, we will confirm with $N$-body integrations that the system is naively unstable on short timescales. 

One way to stabilize these compact systems is to initialize systems in resonant configurations, which can increase the stability time by orders of magnitude \citep{obertas_2017}. In the HIP-41378 system, planets $b$ and $c$ lie close to a $2{:}1$ MMR while the outer three planets lie in or near a $4{:}3-3{:}2$ resonant chain. The fact that we are able to observe HIP-41378 in its present-day configuration points the system being in a resonant architecture. While the proportion of orbital configurations consistent with a resonance are a very small fraction of the parameter space encompassed by the observational constraints, \cite{tamayo2017convergent} showed that slow convergent migration preferentially places planets in these resonant configurations. This process of convergent migration has been invoked to justify the long-term stability of resonant chains such as TRAPPIST-1 \citep{tamayo2017convergent}, TOI-1136 \citep{dai_2023} and HD 110067 \citep{lammers_winn_2024}. 

\subsection{Formation via Convergent Migration}
\label{sec:mig_stab}
\begin{figure}
    \centering
    \includegraphics{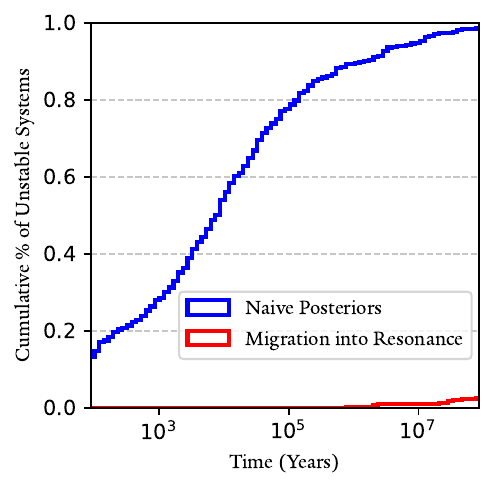}
    \caption{Cumulative distribution of unstable configurations of the HIP-41378 f system as a function of time. In the blue are initial configurations drawn from the posteriors reported by \cite{santerne_2019} and reproduced in Table \ref{tab:params}. In the red are parameters consistent with resonance via convergent migration. We see that the simulations consistent with migration into resonance are overwhelmingly more stable than the naive posteriors. Given the age of the system, we consider this strong evidence that the HIP-41378 system migrated into a resonant chain.}
    \label{fig:stability}
\end{figure}

We use a similar stability analysis to argue that the outer three planets in the HIP-41378 system are indeed in a $4{:}3{-}3{:}2$ resonant chain that was created via convergent migration. We first show that systems not in resonance are overwhelmingly unstable. To do so, we have run a suite of 500 \textit{N}-body simulations. We consider only the outer three planets $d$, $e$ and $f$ -- these are the planets most dynamically relevant to planet $f$, and the distance between planets $c$ and $d$ render the inner two planets dynamically irrelevant to the stability of the outer three. Removing the inner two planets allows us to use a significantly larger timestep in our simulations at great computational gain. We use the \texttt{WHFAST} \citep{rein_2015} integrator in \texttt{REBOUND} \citep{Rein_2012}. We draw all relevant orbital parameters from the posterior distributions given in Table \ref{tab:params}, using $m_d = 4.6 \: M_\mathrm{E}$. We select a timestep equal to $1/15$th of planet $d$'s orbital period, and integrate for $10^8$ years. Integrations were halted if the Hill radii of any pair of planets overlapped, a common metric for a system's instability \citep[e.g.][]{obertas_2017, tamayo_SPOCK}. Our results are shown in the blue curve in Figure \ref{fig:stability}. We see that the vast majority of systems experience rapid instability, with approximately $90\%$ of systems going unstable within 1 Myr. Only 7 systems are stable over the full $10^8$ years.

Next, we explore initial conditions consistent with migration into resonance. We follow the prescription of \cite{tamayo2017convergent}, which has been widely adopted in the literature \citep[e.g.][]{siegel_2021, macdonald_2022, dai_2023, lammers_winn_2024}. Exploiting the scale-free nature of Newtonian gravity, we set the semimajor axis of planet $d$ to $a_\mathrm{d} = 1 \text{ AU}$. We initialize planets $d$, $e$ and $f$ on circular orbits, with consecutive planets initialized $2\%$ wide of their present-day mean-motion resonance. The planetary masses and inclinations are randomly drawn as in the previous runs. We modelled convergent migration using the \texttt{modify\_orbits\_forces} \citep{kostov2016, papaloizou_2000} prescription in \texttt{REBOUNDx} \citep{tamayo2020reboundx} as follows. Exponential semimajor axis damping was applied to planet $f$ only with a timescale $\tau_a = 5 \times 10^6 P_d$. Eccentricity damping was applied to all planets with a timescale of $\tau_e = \tau_a / K$, where $K$ was drawn from a log-uniform distribution $\in \{10, 10^3\}$. Each simulation was integrated for one semimajor axis damping timescale $\tau_a$, upon which damping forces were adiabatically removed over a timescale of $5 \tau_e$. We then discard any simulations that fail to lock into the desired $4{:}3{-}3{:}2$ resonant chain. The successful simulations are rescaled such that $P_\mathrm{d} = 278.36 \text{ days}$, the present-day orbital period of planet $d$, and integrated for $10^8$ more years. At any point, if the mutual separation between any pair of planets is less than the hill radius of the innermost planet or if any planet's semimajor axis exceeds $3$ AU, we consider the system to be unstable and halt the simulation.

We ran $500$ migration simulations. $51$ went unstable during the migration phase, and $1$ failed to lock into the correct resonant chain. The remaining $448$ simulations are shown in the red line in Figure \ref{fig:stability}. The initial conditions consistent with migration into the resonant chain are significantly more stable -- $97\%$ are stable over the full $10^8$ years. Thus, the present-day stability of the system is strong evidence that the HIP-41378 system experienced convergent migration in its primordial history to arrive at the present-day resonant chain.

\section{Transiting Planetary Rings}
\label{sec:rings}
In this section we discuss how an opaque planetary ring system imprints on the transit lightcurve, as a function of its orientation and extent. We show that a realistic ring system around HIP-41378 f is capable of reproducing its transit depth, and in many cases can be detected in transit ingress/egress with the capabilities of JWST.

\subsection{Ring Extent}
\label{sec:ring_extent}

The outer extent of a planetary ring system is governed by the Roche Radius \citep{murray_dermott_2000,schlichting_2011, piro_2020}:
\begin{equation}
\label{eq:roche}
        R_\mathrm{Roche} = 2.45 \left(\frac{3 m_\mathrm{p}}{4 \pi \rho_{\mathrm{ring}}}\right)^{1/3}
\end{equation}
beyond this radius, debris which would make up a ring system will instead coalesce into a moon. The primary degree of freedom in this expression is $\rho_{\mathrm{ring}}$, the density of the ring particles. Assuming zero albedo and full heat redistribution the equilibrium temperature of HIP-41378 f is 294 K \citep{santerne_2019}, which is in excess of the melting point of water ice. The rings around HIP-41378 f therefore must be rocky in composition. 

Estimating a minimum reasonable density for rocky ring particles thus provides a corresponding maximum ring extent. To inform this lower limit we use the work of \cite{babadzhanov_2009}, who find $\rho \sim 0.4$ g/cm$^3$ for the most porous meteorites. Plugging in the most optimistic mass estimate for HIP-41378 f $m_\mathrm{f} = 15 \: M_\mathrm{E}$ and this fiducial lower limit into Equation \eqref{eq:roche}, we see that HIP-41378 f is in principle capable of hosting rings that extend over $14 \: R_\mathrm{E}$, well exceeding the implied $R_\mathrm{f} = 9.2 \: R_\mathrm{E}$ from the transit observations. Thus, it is in principle possible for HIP-41378 f to support a ring system large enough to reproduce the anomalous transit depth observed by \cite{santerne_2019}. This is of course assuming the lowest possible ring particle density and the most optimistic ring configuration of directly face-on.

\subsection{Ring Orientation}
We consider now how the orientation of the ring affects the transit depth. The orbital dynamics of planetary satellites/ring particles is governed by the interplay between the solar tide and the oblateness of the host planet. In the presence of dissipation, satellites, ring particles and circumplanetary disks will damp to the Laplace surface, with characteristic length \citep{tremaine_2009}:

\begin{equation}
    R_\mathrm{L}^5 = J_2 R_p^2 a_p^3 (1 - e_p^2)^{3/2} \frac{m_p}{M_*}
\end{equation}
where $J_2$ is the quadrupole gravitational harmonic, $R_p$ the planet's radius, $a_p, e_p$ the semimajor axis and eccentricity of the planet's orbit around the host star, and $M_*$ the mass of the host star. The orbital inclination of a satellite $i$, or the angle between the satellite's orbit normal and the planet's spin axis, is given \citep{tremaine_2009}:

\begin{equation}
\label{eq:warp}
    \tan 2 i = \frac{\sin 2 \theta}{1 - 2 R_\mathrm{L}^5 / a^5}
\end{equation}
where $a$ is the semimajor axis of the satellite's orbit about the planet, and $\theta$ is the planetary obliquity defined as the misalignment between the planet's orbit normal and its spin axis. Inspection of Equation \eqref{eq:warp} reveals that broadly speaking, the dynamics of a satellite or ring particle orbiting with $a < R_L$ will be dominated by the oblateness of the planet and be coupled to the planet's spin axis. Conversely, the dynamics of a satellite orbiting with $a > R_L$ is dominated by contributions from the host star and will be coupled to the planet's orbital axis. This results in a warped profile for disks that extend past $R_L$. Dynamics of the Laplace surface have been studied in depth by many authors \citep[e.g.][]{zanazzi_lai_2017, speedie_2020, farhat_2021}.

These dynamics are significant for close-in planets, but not HIP-41378 f. For all reasonable values of $J_2$, $R_p$ and $\rho_\mathrm{ring}$, $R_L \gg R_\mathrm{Roche}$, meaning the disk does not extend to the Laplace radius. Hence, to a very good approximation a putative ring system around HIP-41378 f would be completely coupled to its equator. The orientation of the ring system is thus defined completely by the planet's spin vector. The unit spin vector of the planet can be described with two angles: the obliquity $\theta$ defined previously, and the phase angle $\phi$ defined as the projection between the planet's spin axis projected onto the plane of the orbit, and the ascending node of the orbit.

\subsection{Transit Depth}
\begin{figure}
    \centering
    \includegraphics[width=0.5
    \textwidth]{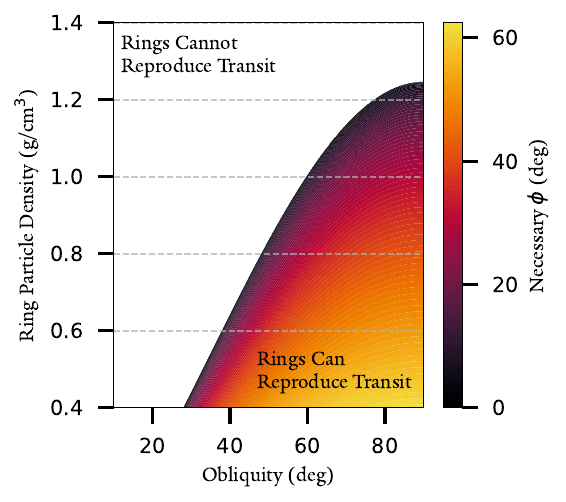}
    \caption{Orientation of the ring system, parameterized by $\phi$, necessary to reproduce the transit depth of HIP-41378 f as a function of planetary obliquity and ring particle density. The white area are areas of parameter space where rings cannot reproduce the transit depth observed even with the most optimistic configuration with $\phi = 0$.}
    \label{fig:ring_extent}
\end{figure}

Consider the standard coordinate system where the $z$-axis points along the line of sight to the system, and the $x$-axis is aligned with the planet's ascending node. The area circumscribed by the outer edge of the ring seen in this frame is its projected area in the $xy$ plane, and is given:

\begin{equation}
\label{eq:ring_area}
\begin{split}
    A_\mathrm{ring} & = R_\mathrm{Roche}^2 \pi \cos \phi \sin \theta \\
    & = \left[ 2.45 \left(\frac{3 m_\mathrm{p}}{4 \pi \rho_\mathrm{ring}}\right)^{1/3}\right]^2 \pi \cos \phi \sin \theta
    \end{split}
\end{equation}
For a given $\theta$ and $\phi$, the transit extent of the ring varies inversely with ring particle density. Figure \ref{fig:ring_extent} shows, for a given planetary obliquity and ring particle density, the ring orientation necessary to reproduce the observed transit depth of HIP-41378 f. The white regions represent areas of parameter space where even for the most optimistic face-on configuration, rings are insufficient. The darkest areas of the colored contours represent the regime where a nearly face-on ring is required, and this orientation requirement is relaxed for lighter colors. We see that there is a plentiful region of parameter space where rings indeed are sufficient to reproduce the transit depth.

\subsection{Detection in ingress/egress}
Detection of planetary rings is in principle observable through various avenues, but mostly notably in artifacts during transit ingress/egress that differ from a purely spherical planet \citep{barnes_fortney_2004, aizawa_2018, akinsanmmi_2018, piro_2020}. This is very similar to the procedure used to measure planetary oblateness \citep{seager_hui_2002, barnes_fortney_2003, zhu_2014, akinsanmi_2024, cassese_2024}. While to date no planetary ring systems have been confirmed via this method, with the advent of JWST these measurements are imminently possible. In this subsection, we briefly demonstrate the feasibility of this method.

\begin{figure}
    \centering
    \includegraphics[width=0.5\textwidth]{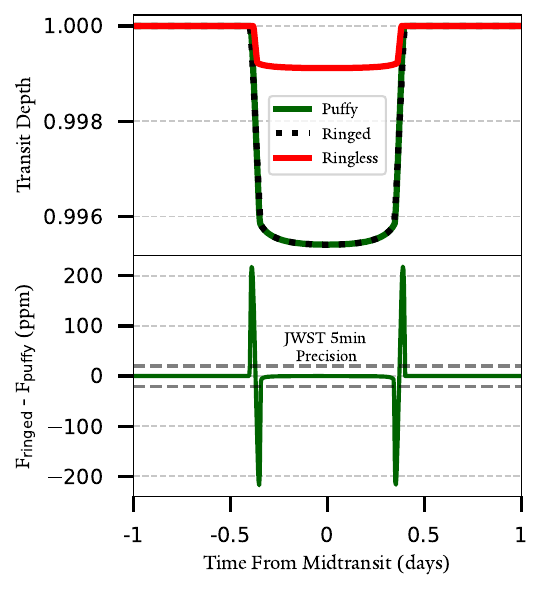}
    \caption{Comparison of transit lightcurves of a flat planet and a planet with an inclined ring system with identical transit depths. The ring system is modelled with $\theta = 45^\circ, \phi = 20^\circ$, and $R_\mathrm{Roche} = 11.29 \: R_\mathrm{E}$. We see that there are large deviations in ingress/egress in which a ringed system can be differentiated from a spherical planet.}
    \label{fig:transit_lightcurve}
\end{figure}
Figure \ref{fig:transit_lightcurve} demonstrates the feasibility of detecting realistic ring systems around HIP-41378 f via transit ingress/egress. We use \texttt{squishyplanet} \citep{cassese_2024} to generate three simple transit lightcurves, which are shown in the top panel. All three lightcurves are generated assuming a circular orbit at $a = 1.37$ AU and quadratic limb darkening parameters $u_1 = 0.0678, u_2 = 0.118$. These system-specific coefficients were derived using the \texttt{ExoTIC-LD} package \citep{grant_2024}, the stellar atmosphere grids from \citet{magic_2015}, and assume observations are collected in the JWST NIRSpec's G395H/F290LP bandpass. The planets are assumed to transit with an impact parameter of $b = 0$. We simulate a perfectly spherical planet with the implied measured radius of $r_\mathrm{f} = 9.2 \: R_\mathrm{E}$ (labelled ``Puffy"), as well as a ring system with $\theta = 45^\circ$, $\phi=20^\circ$ and $R_\mathrm{Roche} = 11.29 \: R_\mathrm{E}$ (labelled ``Ringed") which requires a ring particle density of $\rho = 0.68$ 
g/cm$^3$. The simulated ring system and planet have the same projected area, and at the scale of the first subplot the two curves appear identical. Finally, we also simulate a more realistic spherical HIP-41378 f with a density of $1$ g/cm$^3$ -- this corresponds to a radius of $4.04\: R_\mathrm{E}$, and is labelled ``Ringless". Note the large enhancement in transit depth achieved by a realistic ring system. The lower panel shows the difference in flux between the ``Ringed" lightcurve and the ``Puffy" lightcurve in ppm. Clear deviations in transit ingress/egress are visible. A simple estimate using PandExo \citep{batalha_2017} suggests that JWST's NIRSpec instrument, while operating in its BOTS mode with the G395H grating + F290LP filter, is capable measuring the white light flux with a precision of $< 20$ ppm/hour, which is in principle more than sufficient to detect this deviation and hence differentiate a ringed planet from a puffy planet with equivalent surface area.

We emphasize that these simple comparisons are designed only to demonstrate the feasibility of detecting ring systems. More detailed analysis exists elsewhere in the literature, accounting for factors such as scattered and reflected light as well \citep[e.g.][]{barnes_fortney_2004, sucerquia2020scattered, zuluaga2022bright}. Our model does not account for these factors, nor the gap between the inner edge of the ring system and the outline of the planet. As the primary purpose of this work is to investigate the dynamics of the system, we do not extensively investigate the observational consequences.

\section{Secular Spin-Orbit Resonance}
\label{sec:spin_orbit_resonance}
Secular spin-orbit resonance is a well-studied phenomenon that has been shown to be a plausible origin for the nonzero obliquities of solar system bodies and exoplanets alike. In brief, a planet's obliquity (the angle between its spin axis and orbit normal) may be excited to high values if there is a near match between the precession rates of its spin axis and orbit normal.

Many examples of spin-orbit resonance are present in the solar system. Our moon is perhaps the most prominent example - its $6.68^\circ$ obliquity arises as a consequence of a near match between its orbital and spin precession rates \citep{colombo_1966, peale_1969,touma1998resonances}. In more complex systems such as the solar system, this coupling can occur between the the precession of the spin axis and any of the fundamental frequencies contributing to the nodal precession of the orbit. This has implications for the chaotic obliquity variations of the inner planets \citep{ward_1973, touma_1993, laskar_1993, zeebe_2022, zeebe_2024} as well as the $3^\circ$ degree obliquity of Jupiter \citep{Ward_2006} and the large $89^\circ$ degree obliquity of Uranus \citep{boue_laskar, millholland_batygin_2019, rogoszinski_2020, rogoszinski_2021, lu_2022, saillenfest_uranus}. The most well-studied and accepted case of spin-orbit resonance in our solar system is Saturn, whose $26^\circ$ obliquity is almost certainly due to a near match with a nodal frequency dominated by Neptune \citep{ward_hamilton_2004, hamilton_ward_2004, saillenfest_sat2, saillenfest_sat1, wisdom_2022}.

Naturally exoplanet obliquities are significantly more difficult to detect, though progress has been made on this front - see \cite[Poon et. al. in prep]{bryan_2018, bryan_2020, bryan_2021}. However, there exists a wealth of theoretical literature suggesting that significant exoplanetary obliquities may be common via spin-orbit resonances \citep[e.g.][]{yan_2018, millholland_batygin_2019, quarles_2019, su_lai_2020, li_2021, su_lai_2022, su_lai_2022_se, chen_2023, millholland_2024}.

In this section, we will describe in detail the theory behind secular spin-orbit resonance.

\subsection{Spin Axis Precession}
In the presence of torques from the host star, a planet's spin axis will precess about its orbit normal. The period of precession is given \citep{goldreich1965}:

\begin{equation}
    T_\alpha = \frac{2 \pi}{\alpha \cos \theta}
\end{equation}
where $\theta$ is the planetary obliquity, and is defined by the angle between the planet's orbit normal and its spin axis. The precession rate is defined by $\alpha$, which is primarily a function of the physical parameters the characterize the planet \citep{ward_hamilton_2004, millholland_batygin_2019}

\begin{equation}
    \label{eq:spin_prec}
    \alpha = \frac{3 n^2}{2\Omega} \frac{J_2}{C} = \frac{1}{2} \frac{M_*}{m_p} \left(\frac{R_p}{a}\right)^3\frac{k_2}{C} \Omega
\end{equation}
where $M_*, m_p$ are the masses of the star and planet respectively, $n$ is the orbital mean motion, $\Omega$ the planet's spin frequency, $J_2$ the quadrupole strength of the gravitational field, $R_p$ the planetary radius, $a$ the semimajor axis, $k_2$ the tidal Love number, and $C$ the moment of inertia normalized by $m_p R_p^2$. This precession frequency can be enhanced significantly by the presence of a circumplanetary disk \citep{millholland_batygin_2019} or a moon \citep{saillenfest_uranus, wisdom_2022}.

\subsection{Nodal Precession}
In the presence of torques from the other planets in the system, the planet's orbit normal will precess about the total angular momentum of the system, or the invariant plane normal. The period of nodal precession is given by

\begin{equation}
    T_g = \frac{2 \pi}{|g|}
\end{equation}
where $g$ is the rate of change of the planet's longitude of ascending node. The dynamics governing the rate of nodal precession in a multi-planet system is quite complex and in most cases analysis is performed numerically. However, first-order conclusions and intuition can be drawn from Laplace-Lagrange secular theory \citep{laplace_1775, lagrange_1778, laskar2013solar, murray_dermott_2000}. At this level of approximation, the time evolution of the inclination $I$ and ascending node $\Omega$ of a planet can be calculated as sum of sinuisoidal contributions \citep{ward_hamilton_2004, murray_dermott_2000}:

\begin{equation}
\begin{split}
    \sin \frac{I}{2} \sin \Omega & = \sum_j \frac{I_j}{2} \sin (g_j t + \delta_j) \\
    \sin \frac{I}{2} \cos \Omega & = \sum_j \frac{I_j}{2} \cos (g_j t + \delta_j)
\end{split}
\end{equation}
where $I_j, g_j, \delta_j$ are secular amplitudes, frequencies and phases that depend on the orbital architecture of the system. In a given system there may be many such terms, but for a system with $N$ planets there are typically $N$ large-amplitude terms that end up being significant to the dynamics of the system, with each term dominated -- but not necessarily solely associated with -- one of the planets. Mean motion resonances complicate this picture slightly, but an analytic solution is in principle still possible \citep[e.g.][]{wisdom_1985,malhotra_1989, hadden_2022}. However, in practice these frequencies are often found numerically \citep[e.g.][]{yan_2018, millholland_2024}.

\subsection{Cassini States and Resonance Capture}
Cassini States are equilibrium configurations of the spin axis \citep{colombo_1966, peale_1969, ward_hamilton_2004, fabrycky_2007, su_lai_2020}. They correspond to configurations in which the system's invariant plane normal $\hat{k}$, the planet's orbit normal $\hat{n}$, and the planet's spin axis $\hat{\Omega}$ are coplanar, and $\hat{\Omega}$ and $\hat{n}$ precess about $\hat{k}$ at the same rate. Hence, in a coordinate frame centered on $\hat{n}$ rotating with angular velocity $g$ the Cassini States appear stationary. Defining the planetary obliquity $\theta$ as the angle between $\hat{n}$ and $\hat{\Omega}$, the Cassini State obliquities may be expressed as a function of the ratio $\alpha/g$:

\begin{equation}
    \alpha / g \cos \theta \sin \theta + \sin \left(\theta - I \right) = 0
\end{equation}
where $I$ is the orbital inclination relative to the invariant plane, or the angle between $\hat{n}$ and $\hat{k}$. There are in general $4$ equilibrium solutions, but only Cassini State $2$ is characterized by high $\theta$, which is seen for $\alpha / g > 1$.

Through linearization of the equations of motion that govern the spin axis \citep[e.g.][]{ward_1974, ward_1979}, it can be shown that if $\alpha \cos \theta \sim |g_i|$ for some $i$, the amplitude of the forcing associated with this frequency grows very large and the dynamics of the spin axis are well-approximated by setting $g\sim g_i$. Hence, there exists a high-obliquity Cassini state equilibrium if there is a near match between $\alpha \cos \theta$ and any of the fundamental frequencies $g_i$.

Gaseous planets such as HIP-41378 f are naively expected to form with $\theta \sim 0$ as they accrete gas from the circumstellar disk . Hence, the existence of a high-obliquity equilibrium state is insufficient -- there must be a mechanism to reach it. Capture into spin-orbit resonance is one such mechanism, and occurs when the ratio $\alpha / g$ evolves through unity from below. If this evolution is slow enough to satisfy the adiabatic criterion -- that is, if the timescale of evolution is slow in comparison to the timescale of the spin-axis libration -- phase-space volume will be conserved as the ratio $\alpha / g$ evolves. Hence, a trajectory that starts near Cassini State 2 will remain close to it as the equilibrium point grows in obliquity, and hence the spin vector itself will be excited to high obliquity. This slow change in $\alpha / g$ can be plausibly generated by both evolution in $\alpha$ \citep[e.g.][]{rogoszinski_2020, saillenfest_jupiter, saillenfest_sat1, saillenfest_sat2, wisdom_2022, saillenfest_hip} or $g$ \citep[e.g.][]{millholland2019obliquity, millholland_batygin_2019, lu_2022, millholland_2024}.

\section{The Obliquity of HIP-41378f}
\label{sec:hip_sims}
For a high-obliquity Cassini State 2 to exist for HIP 41378 f, there must be a near match between the spin axis precession rate $\alpha$ and one of the components of the nodal recession $g_i$. In this section, we first assess the likelihood of HIP-41378f presently being in a high-obliquity Cassini State. We then present a migration-driven resonance capture scenario that can plausibly excite the planetary obliquity of HIP-41378 f, and verify with \textit{N}-body simulations.

\subsection{Spin Equilibria}
\label{sec:spin_equilibria}
\begin{figure}
    \centering
    \includegraphics[width=0.5\textwidth]{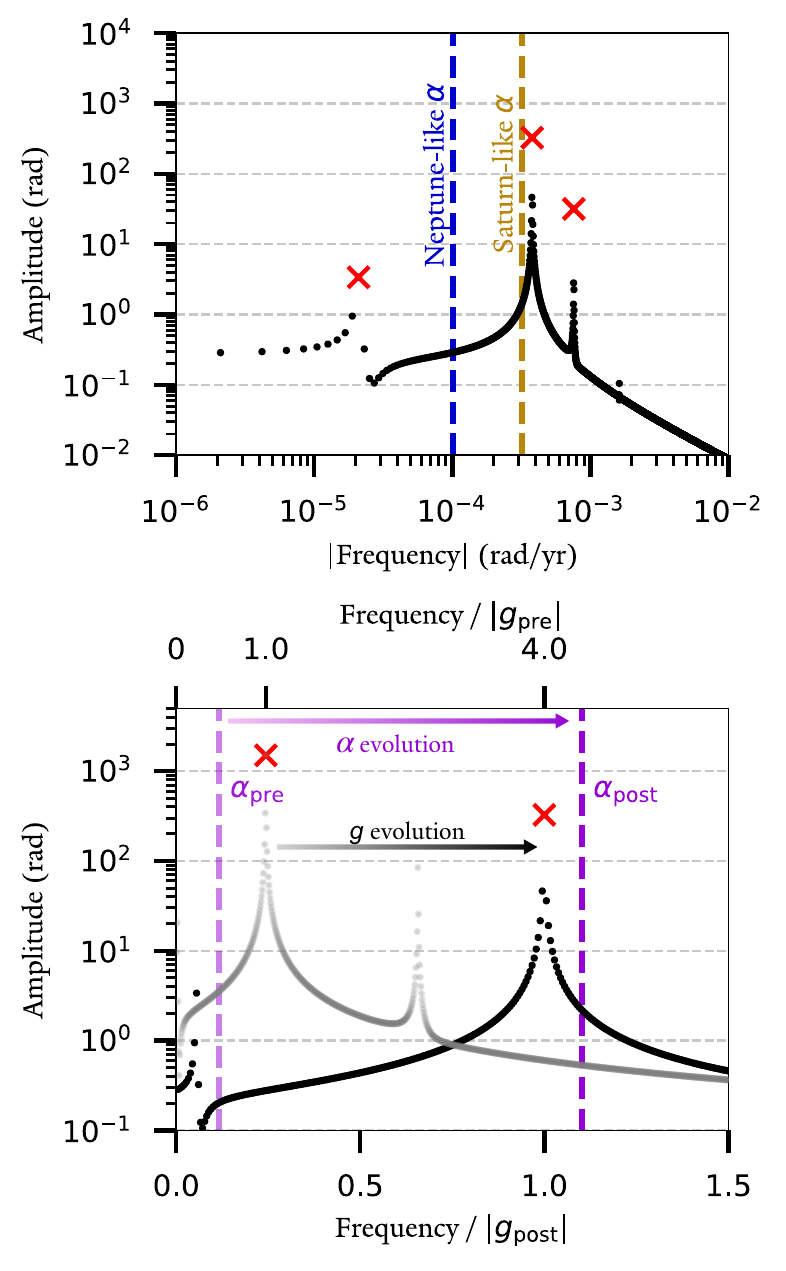}
    \caption{Analysis of precession rates in the HIP-41378 f system, showing that capture into secular spin-orbit resonance is possible for reasonable parameters. The upper subplot shows the present-day power spectrum of the nodal recession of HIP-41378 f in black, with the peaks marked with red Xs. The spin-axis precession rates of a Neptune-like and Saturn-like planet are plotted in the vertical dashed lines, and we see that there is a near-match between the highest-amplitude peak of the nodal recession power spectrum and the Saturn-like spin axis precession rate. The lower panel shows also shows a power spectrum of a fiducial pre-migration system in the gray dots, as well as the spin-axis precession of a Saturn-like planet with slightly enhanced oblateness both pre- and post- migration in the purple vertical dashed lines. We see that $|\alpha / g|$ evolves from below unity pre-migration to above post-migration, satisfying the criteria for resonance capture and subsequent high planetary obliquity.}
    \label{fig:freq_analysis}
\end{figure}

We first obtain the $g_i$ frequencies of the present-day system numerically, following the procedure enumerated in \cite{yan_2018}. First we construct a time-series of the orbital inclination modulated by the longitude of ascending node of planet f, $i(t) e^{\sqrt{-1}\Omega(t)}$. The initial conditions of these simulations are drawn from the set of configurations consistent with convergent migration, from Section \ref{sec:mig_stab}. For each simulation we integrate for $3 \times 10^6$ years, recording outputs every $10$ years. We then perform a Fourier Transform on the resulting time-series data, which is displayed in black dots in the top panel of Figure \ref{fig:freq_analysis}. While in principle the Fourier spectrum depends on the precise orbital configuration of the system and as such will be different for each draw of the system due to the observational uncertainties, in practice these uncertainties are small enough that they do not significantly affect the power spectrum. We have marked each peak of the Fourier spectrum which exceeds an amplitude of unity with red Xs. These are taken to be the fundamental $g$ frequencies of the present-day system. We observe $3$ high-amplitude peaks.

While the nodal recession can be calculated from precise constraints on the orbital architecture, many of the physical parameters necessary to calculate the spin axis precession rate are entirely unconstrained. Thus we use the solar system bodies as fiducial estimates for $\alpha$, which are plotted in the dashed colored lines in Figure \ref{fig:freq_analysis}. The blue and golden dashed lines represent $\alpha$ for an HIP-41378 f at the present-day semimajor axis of $a_f = 1.37$ AU and $C = 0.25$ with Neptune-like and Saturn-like rotation rate and $J_2$, respectively. We see that there is a near-match with highest amplitude peak of the Fourier spectrum with $g_\mathrm{peak} = 3.8 \times 10^{-4}$ rad/year for the Saturn-like case, implying an high-obliquity spin equilibrium exists if HIP-41378 f is around or slightly more oblate than Saturn (the most oblate planet in our solar system). To achieve a near-match with the other two other fundamental frequencies unrealistic values of $\alpha$ are required. For the low-frequency peak we require a $J_2$ equal to $1/5$th of Neptune's while a match with the high-frequency peak requires a $J_2$ equal to $3x$ Saturn's. 

\subsection{Migration-Driven Frequency Evolution}
We next explore the possibility of migration-driven evolution of the precession rates $\alpha$ and $g$ exciting a spin-orbit resonance. We again emphasize the need for the ratio $\alpha / g$ to evolve through unity from below. The bottom subplot of Figure \ref{fig:freq_analysis} depicts an example of the evolution of precession rates under the influence of planetary migration. Due to the nature of the Fourier transformation used to analyze the $g$ frequencies it is impossible to fully track the time-evolution of the $g$ power spectrum through the course of migration. Instead, we take a pre-migration snapshot and a post-migration snapshot of the power spectrum, which is sufficient to show the general evolution trend. The post-migration power spectrum is shown in the black dots, and is identical to the upper subplot. The pre-migration power spectrum is plotted in gray, and is analyzed from a system where planets $b$ and $c$ are initialized in their present-day orbits, planet $d$ at $a_\mathrm{d} = 2.9$ AU (approximately $3.3$x its present-day semimajor axis), and planets $e$ and $f$ two percent wide of their respective present-day mean-motion resonances. Planets $d$, $e$ and $f$ are initialized with circular orbits and mutual inclinations drawn from a Rayleigh distribution centered on $0.5^\circ$. In contrast to the upper subplot, the $x$-axis is now a linear scale and we zoom in near the location of the highest-amplitude peaks. The peak we consider has been marked with a red x. The $x$-axis has also been relabelled for convenience -- the bottom labels are in units of the post-migration peak $g_\mathrm{post} = 3.8 \times 10^{-4}$ rad/year, and the upper labels are in units of the pre-migration peak $g_\mathrm{pre} = 9.3 \times 10^{-5}$. We see that convergent migration acts to shift the entire power spectrum towards the right, corresponding to faster nodal precession. 

We also analyze the evolution of the spin axis precession rate. We consider a planet with the best-fit value for HIP-41378 f's mass $m_\mathrm{f} = 12 \: m_\mathrm{E}$, dimensionless moment of inertia $C = 0.25$. We consider a Saturn-like rotation period and the oblateness to a slightly enhanced value of $J_2 = 1.3 \: J_{2, \mathrm{Saturn}}$. This value is selected as a reasonable illustrative example as to achieve $|\alpha_\mathrm{post} / g| > 1$. The pre-migration spin axis precession rate is plotted with a dotted light purple line, and the post-migration precession rate with a dotted dark purple line. Inward migration similarly pushes the spin axis precession rate to faster values. Due to the strong dependence on semimajor axis evident from inspection of Equation \eqref{eq:spin_prec}, the increase in precession rate is even stronger than the corresponding increase in nodal precession rate. Hence, we see that $|\alpha_\mathrm{pre} / g_\mathrm{pre}| < 1 $ and $|\alpha_\mathrm{post} / g_\mathrm{post}| > 1$, satisfying the resonance capture criteria so long as migration is slow enough.

We conclude this subsection by arguing that assuming a reasonable set of physical parameters for HIP-41378 f, convergent migration is capable of capturing the planet into secular spin-orbit resonance and exciting its planetary obliquity. Hence, migration serves to simultaneously stabilize the system over Gyr timescales as discussed in Section \ref{sec:mig_stab}, and drive the precession frequency evolution necessary to generate the high planetary obliquity needed to reproduce HIP-41378 f's anomalous transit depth with a ring system.

\subsection{Migration Simulations}
We now present a suite of $300$ full \textit{N}-body simulations investigating a high obliquity for HIP 41378 f caused by a spin-orbit resonance generated from primordial convergent migration, accounting for migration and self-consistent spin axis evolution. As mentioned in Section \ref{sec:mig_stab}, the present-day stability of the system heavily implies primordial convergent migration into a resonant chain. We will show in this section that this migration serves as a natural mechanism to evolve the ratio $\alpha/g$ to induce spin-orbit resonance and excite a high obliquity. 

The setup of our simulations closely follows that of \cite{millholland_2024}. We initialize the systems in the pre-migration configurations enumerated in Section \ref{sec:spin_equilibria}. We use the \texttt{WHFAST} integrator \citep{rein_2015} and use a timestep equal to $1/10$th of the orbital period of the innermost planet. We simulated convergent migration with the \texttt{modify\_orbits\_forces} prescription in \texttt{REBOUNDx} \citep{tamayo2020reboundx}. In contrast to our stability simulations, we adopt a more realistic prescription used by \cite{delisle_2017, millholland_2024}. In this prescription, all three outer planets experience semimajor axis damping. The timescale of this damping for the $i$th planet is given

\begin{equation}
    \tau_{a,i} = \tau_0 a_i^\beta
\end{equation}
where $\tau_0 = 1 \times 10^8$ and $\beta = -1.7$. This timescale was selected by slowly increasing until the migration was slow enough to reliably capture planet f into spin-orbit resonance -- as long as the migration timescale exceeds the spin-axis libration timescale the specific choice of migration timescale does not significantly impact our simulations. All three planets also experience eccentricity damping on a timescale $\tau_{e, i} = \tau_{a, i} / 100$. The simulation is integrated until planet d reaches its present-day semimajor axis of $a_d = 0.88$ AU. At this point, all migration forces are turned off and we integrate for an additional $3$ Myr before halting the simulation.

We account for self-consistent spin and dynamical evolution using the prescription of \cite{eggleton_1998} and \cite{mardling_lin_2002}, implemented by \cite{Lu_2023} in \texttt{REBOUNDx}. We endow planet $f$ with structure and approximate the other four planets as point particles. The relevant additional parameters needed to describe the spin evolution of the planet are the initial direction and magnitude of the spin axis $\Omega_\mathrm{f}$, the radius of the planet $r_f$, the tidal Love number $k_2$, and the dimensionless moment of inertia $C$. Note that tides are not expected to be important for the spin-axis evolution of this planet, so we do not include their effects in the interest of minimizing computation time. We vary the density of HIP-41378 f from $\rho_\mathrm{f} \in \{0.7, 1.5\}$ g/cm$^3$ and $J_2 \in \{4.75 \times 10^{-3}, 7.86 \times 10^{-1}\}$ -- for comparison, Saturn's $J_2$ moment is $1.65 \times 10^{-2}$, so our simulations range from $30\%$ to $48$x Saturn's $J_2$ moment. These initial conditions are somewhat arbitrarily selected, but are designed to more than encompass a range of realistic physical parameters that would result in a near-match with the high-amplitude frequency peak identified in Section \ref{sec:spin_equilibria}.

\begin{figure*}
    \centering
    \includegraphics{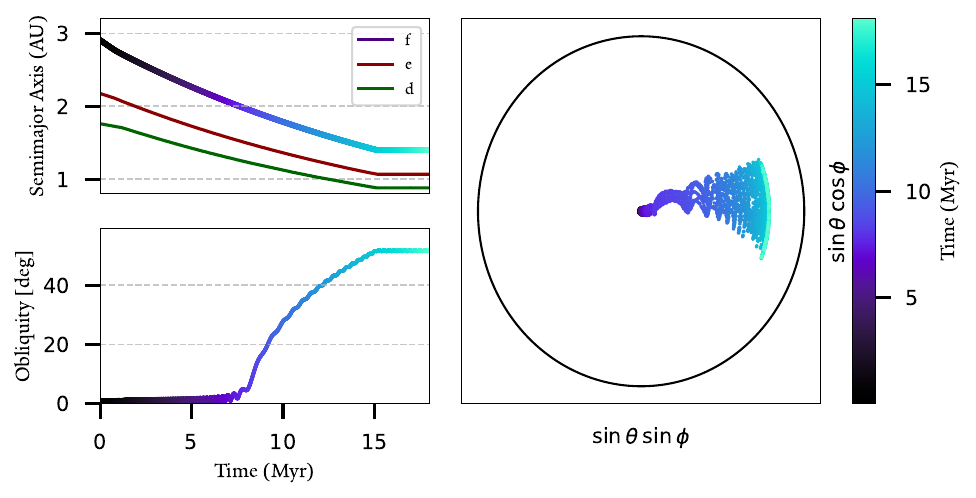}
    \caption{A fiducial case of the dynamical evolution of the HIP-41378 system, which leads to secular spin-orbit resonance and high planetary obliquity. The upper-left subplot shows convergent migration of the three outer planets. Each is initialized just wide of their present-day mean-motion resonances and are quickly caught into them, and migrate inward in lockstep afterwards. At around $t{\sim} 15$ Myr, we turn the migration force off. The bottom left plot shows the planetary obliquity evolution of HIP-41378 f. At around $t{\sim} 8$ Myr, the spin-orbit resonance is entered and the planetary obliquity is steadily excited until migration is turned off. The obliquity is stable at $52^\circ$. The right hand plot shows a polar view of the evolution of the planetary spin axis in a frame that precesses with the planet's orbit. The distinctive banana-like librations of capture into a Cassini State are visible.}
    \label{fig:convergent_migration}
\end{figure*}

Figure \ref{fig:convergent_migration} shows one of our simulations in detail. This simulation is initialized with $J_2 = 0.06641$ and $\Omega_\mathrm{f} = 5$ hours. All three planets migrate inward and are caught into their respective mean-motion resonances at around $t \sim 1$ Myr, at which point they migrate inward in lockstep. At around $t = 8$ Myr, the spin-orbit resonance is reached and the planetary obliquity slowly grows until the migration is halted at $t \sim 15$ Myr, at which point it remains stable with small oscillations due to libration around the fixed point at around $50^\circ$.

\begin{figure}
    \centering
    \includegraphics{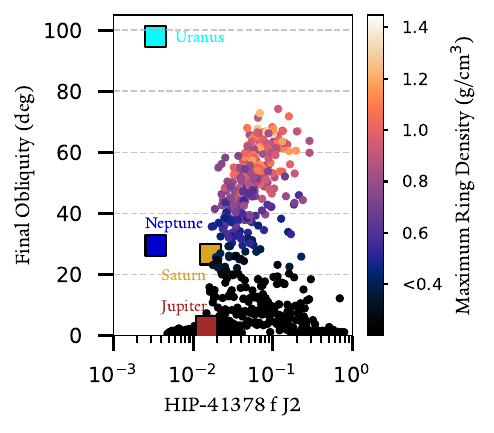}
    \caption{Population-level statistics of our migration simulations, plotting final planetary obliquity vs the $J_2$ moment of planet f. The colorbar represents the density of ring particles needed to generate a large enough ring consistent with the observed transit depth for the given final obliquity, assuming the most optimistic face-in geometry. The black points are incapable of generating the observed transit. For reference the obliquities and $J_2$ moments of the solar system giant planets are also plotted. We see that starting from simulations run with Saturn's $J_2$, planetary obliquities capable of hosting a ring system that generates HIP-41378 f's transit profile are possible.}
    \label{fig:ensemble}
\end{figure}

We also report results from our entire ensemble of simulations, shown in Figure \ref{fig:ensemble}. We plot the final obliquity of planet f against its final $J_2$ moment for each simulation. We also ran similar suites of simulations slightly varying mutual inclination and migration speed, with no significant differences -- the final planetary obliquity is mostly sensitive to the $J_2$ moment of HIP-41378 f. Note that the planet's spin rate does not meaningfully evolve over the course of the simulations, so this $J_2$ is essentially the primordial value. For reference, we have also plotted the obliquities and $J_2$ moments of the solar system giant planets. As predicted from our frequency analysis, starting at $J_2$ slightly higher than Saturn's HIP-41378 f is able enter the spin-orbit resonance and excite high planetary obliquity. The range in $J_2$ for capture into the spin-orbit resonance is approximately Saturn's $J_2$ to around $20$x Saturn's $J_2$ -- planets that are more or less oblate than this range fail to lock into the relevant frequency peak. Not all of our simulations in this range are able to attain high obliquity, which we attribute to the probabilistic nature of resonance capture \citep{su_lai_2022}. The plot is color-coded with the ring particle density needed to generate a ring system with sufficient extent to reproduce the observed transit depth, assuming an orientation with $(\phi=0)$ and the final planetary obliquity. Points in black are unable to host ring systems with sufficient extent without resorting to ring particles more porous than the fiducial limit we discussed in Section \ref{sec:ring_extent}, while the colored points are capable and heavily populate the region of $J_2$ parameter space where spin-orbit resonance is achieved.

Our results indicate that there is a large region of reasonable parameter space in which the convergent migration process can be expected to generate a high enough planetary obliquity in HIP-41378 f to sufficiently warp an opaque ring system out of the orbital plane to masquerade as an unusually puffy planet. We predict that if HIP-41378 f is at least as effectively oblate as Saturn then high planetary obliquities can very reasonably be attained. This assumption is not a given, but is not unrealistic. We briefly enumerate some caveats and limitations of our study. First, the evolution of planetary structure over the course of the simulation is not considered \citep[e.g.][]{lu_2024_hatp11}. The most significant effect overlooked is the evolution of the planetary spin rate, which is not dynamically impacted by the migration process and thus remains constant in our simulations. However, as gas giant planets accrete from the circumstellar disk their spin rates are expected to form near their breakup spin rates, and then magnetohydrodynamical effects work to expel angular momentum from the system and drive the spin rates to the significantly sub-critical values we observe in the giant planets of our solar system \citep{batygin_2018}. This does not change the limits on the range of viable planetary $J_2$s that can excite obliquity, however -- this limit is set by a match between the spin axis precession rate and the present-day orbital frequencies, the latter of which is well-constrained from the orbital solutions. We also did not explore a wide range of initial migration configurations, opting to arbitrarily initialize the planets at roughly twice their present-day orbits. If the migration process is significantly shorter than this, it will be more difficult to induce the requisite $|\alpha/g|$ crossing as the frequencies will evolve less. The initial formation locations of these planets are unconstrained. Finally, in our simulations we did not consider the effect of the ring on the dynamics of the spin axis itself. The only effect the ring is expected to have is a minute enhancement in $\alpha$, which would be far outweighed by the uncertainties in the physical parameters of the planet.

\subsection{The Effect of a Massive Moon}
Our population-level results in Figure \ref{fig:ensemble} predict high obliquities if a certain $J_2$ threshold is achieved. We have discussed $J_2$ thus far in the context of planetary oblateness only, but there are in fact many ways to increase the effective $J_2$ of the planet. One reasonable way is to include the effect of a massive moon, which were shown to be tidally stable around HIP-41378 f by \cite{harada_2023}. In this subsection we briefly describe how a massive moon could function as a form of precession enhancement, which would allow less oblate Neptune-like planets to be caught into spin-orbit resonance.

\begin{figure}
    \centering
    \includegraphics[width=0.5\textwidth]{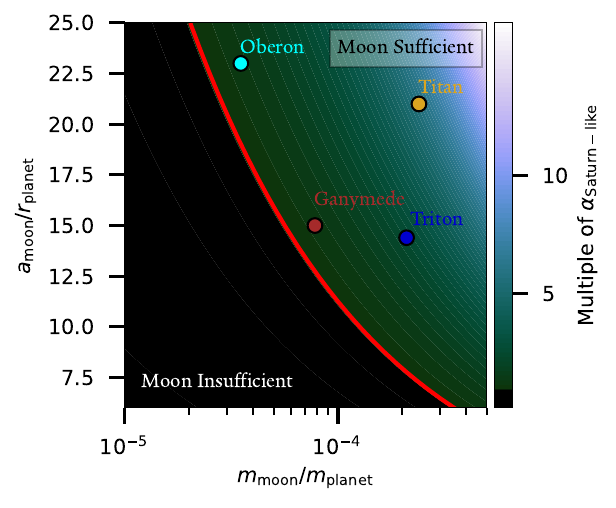}
    \caption{Exploration of how a massive moon enhances spin precession rate. We plot moon mass on the $x$-axis and moon semimajor axis on the $y$-axis, and the colorbar represents the spin precession rate of a planet with Neptune-like oblateness, spin rate and moment of inertia at the present-day orbit of HIP0-41378 f, enhanced by a moon of corresponding mass and orbit. The red line delineates the spin precession rate of a Saturn-like planet at the same orbit, the nominal requirement for capture into spin-orbit resonance as seen in Figure \ref{fig:ensemble}. The black region is parameter space in which the moon cannot sufficiently generate the requisite spin precession rate, with the opposite for the colored region. For reference the mass and semimajor axis ratios of the outermost major satellite of the four solar system giant planets are also plotted, and all are in the allowed parameter space.}
    \label{fig:moon_enhancement}
\end{figure}

We wish to investigate the effect a massive satellite has on the rate of spin-axis precession, given by Equation \eqref{eq:spin_prec}. There are two effects to consider: there are enhancements in $J_2$ moment and the normalized moment of inertia $C$. These can be written \citep{tremaine_1991, ward_hamilton_2004, lu_2022}:

\begin{equation}
\begin{split}
        J_{2, \mathrm{effective}} & = J_{2, \mathrm{planet}} + \frac{1}{2}\sum_i \left(\frac{m_i}{m_p}\right) \left(\frac{a_i}{R_p}\right)^2 \\
        C_\mathrm{effective} & = C_\mathrm{planet} + \frac{1}{m_p R_p \Omega^2} \sum_i m_i a_i^2 n_i
\end{split}
\end{equation}
where $m_i$ is the mass, $a_i$ is the semimajor axis and $n_i$ is the orbital mean motion of the $i$th satellite in the system. Figure \ref{fig:moon_enhancement} displays the precession enhancement a realistic moon may have. The colorbar corresponds to the precession rate $\alpha$ a Neptune-like planet at HIP-41378 f's present-day orbit would exhibit if a moon of a given mass and semimajor axis was orbiting it. The red contour corresponds to the precession rate of a Saturn-like planet, the nominal cutoff for capture into secular spin-orbit resonance in our simulations, in the same location. For reference, the outermost major satellites of the four solar system planets are also plotted. We see that all four satellites are in the region of parameter space capable of sufficiently enhancing the precession rate of a Neptune-like planet to Saturn-like levels. Hence, a realistic satellite greatly expands the allowable physical parameter space of HIP-41378 f that results in capture into spin-orbit resonance. 

We conclude that accounting for the existence of a realistic massive moon essentially allows all reasonable physical parameters associated with HIP-41378 f to result in capture into spin-orbit resonance. A large migrating exomoon was posited as the both the source of HIP-41378 f's obliquity and the ring system itself by \cite{saillenfest_hip} -- in fact, the two scenarios are not incompatible with one another as a migrating exomoon acts to increase the spin precession rate. The detection of exomoon may be feasible in the near future \citep[e.g.][]{kipping_2022}, which would represent additional credence to our theory.

\section{Other Super-Puffs}
\label{sec:others}
We now briefly analyze the possibility of secular spin-orbit resonance in other super-puff systems. While the exoring hypothesis is less necessary to explain these super-puffs due to their closer-in orbits, it remains a viable solution to some of the super-puffs at farther distances \citep{piro_2020}. In addition, for the closest super-puffs tidal heating may render obliquity tides significant \citep{millholland_2019_tides, millholland_2020}. Both of these hypotheses support the intriguing results of \cite{millholland2019obliquity} and \cite{leleu_2024} who showed that sub-Neptunes near resonance tend to be puffier.

Secular spin-orbit resonance requires nodal precession of the orbit, as described in Section \ref{sec:spin_orbit_resonance}. One way to drive nodal precession is the $J_2$ moment of the host star \citep[e.g.][]{brefka_2021, faridani_2023}. However, the most natural way to maintain spin-orbit resonances is companion planets. We thus restrict our attention to super-puffs in multi-planet systems. We thus consider five additional systems: Kepler-223 \citep{mills_2016}, Kepler-177 \citep{vissagragada_2020}, Kepler-359 \citep{hadden_lithwick_2017}, Kepler-51 \citep{masuda_2014}, and K2-24 \citep{petigura_2018}. Intriguingly, a number of these systems also lie in resonant configurations, potential evidence of migration in the system's history \citep[e.g.][]{lee_chiang_2016}. 

%\begin{figure}
%    \centering
    %\includegraphics[width=0.5\textwidth]{architectures.pdf}
    %\caption{System architectures of six multi-planet super-puff systems, along with our solar system for scale. Semimajor axis is plotted on a log scale on the x-axis. Each planet's size corresponds to its mass, and the colorbar to its implied density -- the super-puffs in each system are pink.}
    %\label{fig:puff_architectures}
%\end{figure}

\begin{figure}
    \centering
    \includegraphics[width=0.5\textwidth]{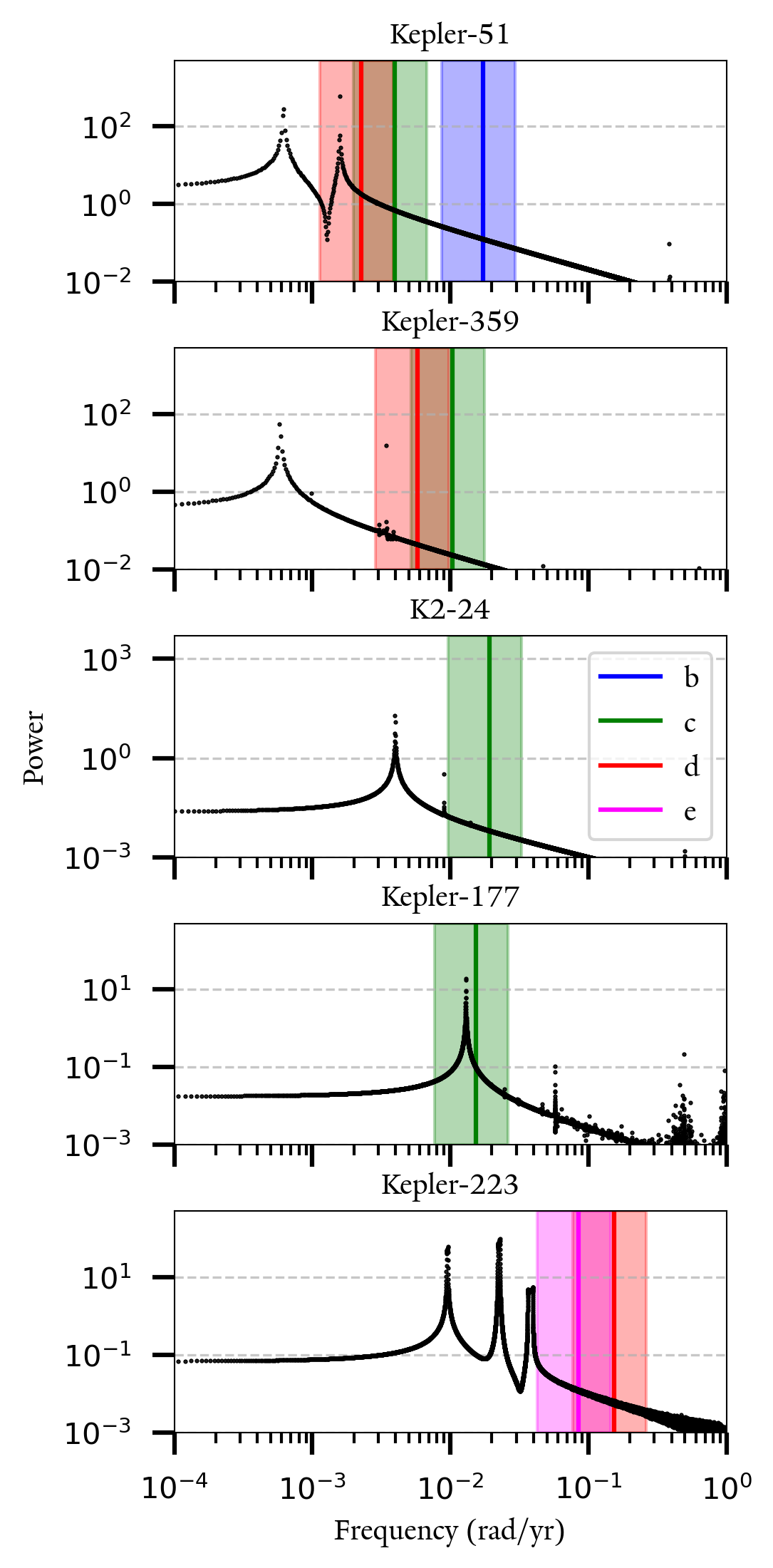}
    \caption{Frequency analysis of five multi-planet super-puff systems. Each subplot is analogous to the upper panel in Figure \ref{fig:freq_analysis}. The vertical colored bands represent reasonable present-day ranges of $\alpha$ for each of the super-puffs in the system, color-coded by designation. We find a near match for at least one super-puff in each system.}
    \label{fig:puff_systems}
\end{figure}

We numerically analyze the frequency power spectra of each system as in Section \ref{sec:hip_sims}. We compare the peaks of this frequency analysis to a set of physically reasonable $\alpha$ values for each super-puff, given by a Neptune-like oblateness and rotation rates ranging fro 10 hours to 36 hours for reference. These are shown in Figure \ref{fig:puff_systems}. In each subplot, the black dots are a representative frequency power spectrum for the system, and the vertical bands correspond to the reasonable $\alpha$ values we explored where each color corresponds to a super-puff in the system. Note that while some of these planets may be expected to be tidally locked, we did not find any near-matches for $\alpha$ precession rates associated with tidally locked states and thus do not depict them. We find possible matches for the following super-puffs: Kepler-51 d, Kepler-359 d, K2-24 c, Kepler-177 c, Kepler-223 e -- in other words, there is a super-puff in every system which can potentially be in a spin-orbit equilibrium.

We briefly comment on the Kepler-51 system, a well-studied system that is in a 1:2:3 resonant chain \citep{masuda_2014} and believed to have arrived at its present-day configuration via convergent migration \citep{lee_chiang_2016}. The recent work of \cite{lammers_winn_k51}, who inferred an unusually slow rotation period of $>40$ hours for Kepler-51 d via a measurement of the (lack of) planetary oblateness. One explanation they discussed was a high planetary obliquity, with $\theta \geq 75^\circ$, which could mask the true oblateness of the planet in the sky-projected obliquity. Their results, taken in conjunction with the near-match in frequency space for Kepler-51 d shown in Figure \ref{fig:puff_systems} and the potential migration history of the system, indicates a high-obliquity state for Kepler-51 d is a very reasonable prospect. We defer more detailed analysis of this, as well as the other super-puff systems enumerated, to future work.

\section{Conclusion}
\label{sec:conclusion}
In this work, we have investigated the dynamical history of the HIP-41378 system, motivated by the anomalous low density of the outermost planet HIP-41378 f. We find strong evidence that the system formed via convergent migration, as the outer three planets lie near a 4:3:2 resonant chain that is dynamically unstable on short timescales otherwise. We also find that if HIP-41378 f has a $J_2$ moment slightly greater than that of Saturn or is accompanied by a massive satellite system, this migration process likely results in capture into secular spin-orbit resonance and significant excitation of the planetary obliquity. If there is an opaque system of planetary rings around HIP-41378 f, this obliquity is in many cases able to reproduce its anomalous transit signal. Hence, we assert that a dynamical history involving convergent migration simultaneously explains both the system's long-term stability as well as in many cases generating a high planetary obliquity. This high planetary obliquity lends credence to the popular theory that HIP-41378 f is not in fact an extremely low-density planet far from its host star, but rather hosts a system of opaque rings \citep{akinsanmi_2020}. We have shown that the spin-orbit configurations arising from convergent migration naturally lead to systems which can host large enough ring systems to reproduce the observed transit depth and anomalous density of HIP-41378 f. We also briefly comment on other super-puffs in multi-planet systems, and show that many of them are also plausibly in high-obliquity states.

We therefore encourage immediate additional observations targeted at HIP-41378 f to verify the true nature of the exoring hypothesis. Our work has demonstrated the first-order feasibility of differentiating between planetary rings and a puffy planet in a transit lightcurve. More nuanced modelling is certainly possible to take advantage of the vast capabilities of JWST, including accounting for the oblateness of the planet itself and rings which are not fully opaque. Scattered \citep{barnes_fortney_2004} and reflected light \citep{arnold_2004} from ring particles may also imprint themselves on the transit lightcurve. Spectroscopic effects \citep{ohta_2009} may also be relevant, among a host of other less obvious effects \citep{heller_2018}.

Super-puffs represent one of the most intriguing unsolved mysteries in exoplanet science today. Our work has highlighted the importance of dynamics in this conversation, which until recently has been primarily a structural debate. While exorings may not be a necessary or even viable explanation for other super-puffs, planetary obliquity may still be highly relevant in the form of obliquity tides. The signatures of tidal heating are readily visible in transmission spectra \citep[e.g.][]{welbanks_2024}. As most super-puffs are observed to be on circular orbits, so hence unable to be heated via eccentricity tides, tidal heating signatures would almost certainly point to significant planetary obliquity. We thus again encourage further atmospheric observations of these super-puffs as potential tests of tidal heating and signs of planetary obliquity.

\bibliography{sample631}{}
\bibliographystyle{aasjournal}
%\allauthors

%\listofchanges

\end{document}